\documentstyle[psfig,12pt]{article}

\def\beq{\begin{equation}}
\def\eeq{\end{equation}}

\def\be{\begin{equation}}
\def\bea{\begin{eqnarray}}
\def\ee{\end{equation}}
\def\eea{\end{eqnarray}}
\def\d{\partial}
\def\eqref#1{(\ref{#1})}

\def\e{{\rm e}}
\def\tr{{\rm tr}}

\def\g{\gamma}

\begin{document}


\begin{titlepage}

\begin{centering}

\vspace*{3cm}

{\Large\bf A solution to the fermion doubling problem for
supersymmetric theories on the transverse lattice}

\vspace*{1.5cm}

{\bf  Motomichi Harada and Stephen Pinsky}
\vspace*{0.5cm}

{\sl Department of Physics \\
Ohio State University\\
Columbus OH 43210}

\vspace*{0.5cm}

\vspace*{1cm}


\vspace*{1cm}

\vspace*{1cm}

\begin{abstract}
Species doubling is a problem that infects most numerical methods
that use a spatial lattice. An understanding of species doubling can
be found in the Nielsen-Ninomiya theorem which gives a set of
conditions that require species doubling. The transverse lattice
approach to solving field theories, which has at least one spatial
lattice, fails one of the conditions of the Nielsen-Ninomiya theorem
nevertheless one still finds species doubling for the standard
Lagrangian formulation of the  transverse lattice.  We will show
that the Supersymmetric Discrete Light Cone Quantization (SDLCQ)
 formulation of the transverse lattice does not have species
doubling.
\end{abstract}
\end{centering}

\vfill

\end{titlepage}
\newpage
\section{Introduction}
When one formulates a theory with chiral fermions on a spatial
lattice, one of the most notorious obstacles is the Nielsen-Ninomiya
theorem\cite{Nielsen:1980rz} which gives a set of conditions that
require species doubling. In our transverse lattice formulation of
field theory we use both a spatial lattice and a momentum lattice.
The transverse lattice formulations usually has some non-local
interaction(s) which voids the Nielsen-Ninomiya theorem however it
still seems to have the species doubling problem
\cite{Burkardt:1998ws}.

Recently, the authors proposed a super Yang-Mills (SYM) model in 2+1
dimensions on a transverse lattice with one exact supersymmetry
\cite{Harada:2003bs}. It is well known  that in the standard
Lagrangian formulation of SYM on the transverse lattice one finds a
fermion species doubling problem. We will show however that we are
free from species doubling when one uses Supersymmetric Discrete
Light Cone Quantization (SDLCQ). This is yet another demonstration
of value of maintaining an exact supersymmetry in the numerical
approximation. Of course two popular methods of dealing with the
doubling, staggered fermions \cite{Susskind:1976jm} and the Wilson
term \cite{Wilson:1974sk}, work for the lagrangian formulation of
SYM theories on a transverse lattice. In addition Chakrabarti, De
and Harindranath recently proposed the use of the forward and
backward derivatives to remove the species doubling on the light
front transverse lattice \cite{Chakrabarti:2002yu}. However those
methods badly break the supersymmetry and it is unclear how many of
the unique properties of supersymmetry persist. While our approach
can only be used for the transverse lattice formulation of
supersymmetric theories, it resolves the doubling problem
automatically.

This paper is organized as follows. In Section~2 we will see that
the species doubling arises in the standard Lagrangian formulation
of the transverse lattice, but can be resolved when one applies the
method proposed by Ref.~\cite{Chakrabarti:2002yu}. In Section~3 we
show that in the SDLCQ formulation of the transverse lattice we do
not have any species doubling. In section~4 we discuss some general
reasons for this result and give the generalization to 3+1
dimensions.

\section{Fermion species doubling problem on a transverse lattice}
To focus on the fermion species doubling problem of the transverse
lattice~\cite{Harada:2003bs}, let us consider fermion fields only by
setting the coupling $g=0$ and the link variables $M,M^{\dag}=1$.
For this theory one spatial dimension is discretized on a spatial
lattice. We work in the light cone coordinates so that
$x^{\pm}\equiv(x^0\pm x^1)/\sqrt 2$ with $x^{\pm}=x_{\mp}$ and
$x^{\perp}\equiv x^2=-x_{2}$ is the dimension that is discretized on
the spatial lattice. The Lagrangian is given by
\[
    {\cal L} =\sum_i\tr\left[\bar{\Psi}_i\g^{\mu}\d_{\mu}\Psi_i+\frac i{2a}
    \bar{\Psi}_{i}\g^{\perp}(\Psi_{i+1}-\Psi_{i-1})\right],
\]
where $i$ is the site index, the trace has been taken with respect
to the color indices, $\mu=\pm$, and $a$ is the lattice spacing. The
gamma matrices are defined to be $\gamma^0=\sigma^2$,
$\gamma^1=i\sigma_1$, and $\gamma^{\perp}=i\sigma^3$ with
$\g^{\pm}\equiv (\g^0\pm \g^1)/\sqrt 2$. For $\Psi_i=2^{-1/4}\left(
\begin{array}{c} \psi_i
\\ \chi_i \end{array} \right)$ we find the equation of motion
\[ \d_-\chi_i=\frac 1{2\sqrt 2 a}(\psi_{i+1}-\psi_{i-1}). \]
Inverting the light cone spatial derivative, we eliminate the
non-dynamical field $\chi_i$ from ${\cal L}$ and get
\[
    {\cal L}= \sum_i\tr\left[i\psi_i\d_+\psi_i+\frac i{8a^2}
    (\psi_{i+1}-\psi_{i-1})\d^{-1}_-(\psi_{i+1}-\psi_{i-1})\right].
\]
Note that the second term is non-local. This is sufficient to avoid
the Nielsen-Ninomiya theorem. The equation of motion for $\psi_i$ is
\begin{equation}
\d_+\psi_i=\frac 1{8a^2}\d^{-1}_-
    (\psi_{i+2}-2\psi_i+\psi_{i-2}). \label{eom}
\end{equation}
We substitute the Fourier transformed form of $\psi_i$,
\[\psi_j(x)=\int_{-\pi/a}^{\pi/a}dk^{\perp}\int_0^{\infty}dk^+dk^-
    \e^{i(k^+x^-+k^-x^+-k^{\perp}(aj))}\tilde{\psi}_j(k),\]
into Eq.~\eqref{eom} to find a dispersion relation
\begin{equation}
k^-=\frac 1{2k^+}\left(\frac {\sin k^{\perp}a}{a}\right)^2.
\label{dispersion}
\end{equation}
Clearly, in the continuum limit where $a\to 0$, we find finite
energy not only at $k^{\perp}\approx 0$, but also at $k^{\perp}
\approx \pm \pi/a$ for $-\pi/a<k^{\perp}<\pi/a$, yielding {\it
extra} unwanted fermion species, that is, the notorious fermion
species doubling problem.

Let us point out that the same equation of motion and thus the same
dispersion relation follow if one uses Heisenberg equation of motion
$i\d_+\psi_{i,rs}(x)=[\psi_{i,sr}(x),P^-]$. This is the approach we
will use in the next section.  In this calculation we use the equal
(light cone) time anticommutation relation
$\{\psi_{i,rs}(x^-),\psi_{j,pq}(y^-)\}=\delta(x^--y^-)
\delta_{ij}\delta_{rp}\delta_{sq}/2a$, where we've explicitly
written out the color indices $r,s,p,q$ and
\[
    P^-\equiv a\sum_i \int dx^-T^{+-}
    =a\sum_i \int dx^-\tr\left[-\frac i{8a^2}(\psi_{i+1}-\psi_{i-1})
    \d^{-1}_-(\psi_{i+1}-\psi_{i-1})\right].
\]
$T^{\mu\nu}$ is the stress-energy tensor.

One might wonder what happens if we tried another difference
operator, for instance, the forward/backward derivative in place of
the symmetric derivative. Answering this question is instructive
since the authors of Ref.~\cite{Chakrabarti:2002yu} have found no
fermion doubling for chiral fermions if one uses forward and
backward derivatives on the light front transverse lattice.
Following their procedure, we get in terms of $\psi_i$ and $\chi_i$
\begin{eqnarray*}
&&  {\cal L} =\sum_i\tr\left[i\psi_i\d_+\psi_i+i\chi_i\d_-\chi_i
    -\frac i{\sqrt 2 a}\left(\chi_i(\psi_{i+1}-\psi_i)
    +\psi_i(\chi_i-\chi_{i-1})\right)\right]\\
&&  =\sum_i\tr\left[i\psi_i\d_+\psi_i+i\chi_i\d_-\chi_i
    -\frac {\sqrt 2 i}{ a}\chi_i(\psi_{i+1}-\psi_i)\right].
\end{eqnarray*}
This yields
\[ \d_-\chi_i=\frac 1{\sqrt 2 a}(\psi_{i+1}-\psi_{i}) \]
and
\[
    {\cal L}= \sum_i\tr\left[i\psi_i\d_+\psi_i+\frac i{2a^2}
    (\psi_{i+1}-\psi_{i})\d^{-1}_-(\psi_{i+1}-\psi_{i})\right].
\]
From this we find a dispersion relation
\[
k^-=\frac 1{2k^+}\left(\frac {\sin
\frac{k^{\perp}a}2}{a/2}\right)^2.
\]
In the continuum limit we find a finite energy only at
$k^{\perp}\approx 0$, meaning that we do not have the doubling
problem. Hence, we found that the method to remove the doubling
proposed in Ref.~\cite{Chakrabarti:2002yu} works even for adjoint
fermions.

\section{Transverse lattice with SDLCQ}
In Ref.~\cite{Harada:2003bs} we proposed a discrete transverse
lattice formulation of the supercharge $Q^-$, which gives the
correct continuum form and the $P^-$ obtained from SUSY algebra
$\{Q^-,Q^-\}=2\sqrt 2P^-$ also gives the correct continuum form.
With this $P^-$ in hand, following the same procedure we did in the
previous section, we set $g=0$ and $M,M^{\dag}=1$ to see whether we
suffer from the fermion doubling problem. This $P^-$ is given by
 \[ P^-=a\sum\int dx^-\tr\left[-\frac i{2a^2}(\psi_{i+1}-\psi_{i})
    \d^{-1}_-(\psi_{i+1}-\psi_{i})\right]. \]
Heisenberg equation of motion yields
\[ i\d_+\psi_{i,rs} =[\psi_{i,sr},P^-]=\frac i{2a^2}\d^{-1}_-
    (\psi_{i+1}-2\psi_i+\psi_{i-1})_{rs}.\]
Hence, it follows that
\[ k^-=\frac 1{2k^+}\left(\frac{\sin\frac{k^{\perp}a}{2}}{a/2}
\right)^2 .\]%
Notice, remarkably, that we have a finite energy {\it only} at
$k^{\perp}\approx 0$, so that we are {\it free from the species
doubling problem with SDLCQ}.

A word of caution is due here. This $P^-$ happens to be the same as
the one obtained in Ref.~\cite{Chakrabarti:2002yu}, where the
authors used the forward and backward derivatives however we get
$P^-$ in a completely different way.

\section{Discussion}
We reviewed the known result that one suffers from a species
doubling problem in the transverse lattice Lagrangian formalism with
the symmetric derivative in spite of the fact that our adjoint
fermions interact non-locally. We applied the method of removing the
doubling proposed by the authors of Ref.~\cite{Chakrabarti:2002yu}
originally for chiral fermions and found that it works as well even
for adjoint fermions. We then showed that we do not suffer from
species doubling in the SDLCQ formulation of the transverse
lattice~\cite{Harada:2003bs}.

While we did the calculation in 2+1 dimensions, we should note that
this doubling persists in 3+1 dimensions. The authors have been
working on the extension of the model in Ref.~\cite{Harada:2003bs}
to a 3+1 dimensional model with transverse lattices in two spatial
directions. We have found \cite{Preparation} that the standard
transverse Lagrangian formulation leads to the following dispersion
relation,
\[ k^-=\frac 1{2k^+}\left[ \left(\frac{\sin k^{\perp}_1a}{a}
\right)^2+\left(\frac{\sin k^{\perp}_2a}{a} \right)^2\right],
\]
where $k^{\perp}_i$ is the $i$-th transverse momentum. For a
model with SDLCQ formulation of the transverse lattice,
\[ k^-=\frac 1{2k^+}\left[ \left(\frac{\sin\frac{k^{\perp}_1a}{2}}{a/2}
\right)^2+\left(\frac{\sin\frac{k^{\perp}_2a}{2}}{a/2}
\right)^2\right].
\]
Again, we do not have any species doubling with SDLCQ.

In Ref.~\cite{Harada:2003bs} we found that the color of physical
states must be contracted at each site. However, this constraint was
derived in the standard Lagrangian formalism, which suffers from the
doubling problem. Therefore, one might ask if there is any change in
the physical constraint due to the doubling problem. We believe the
answer is no. The reason is the following. The physical constraint
we found in \cite{Harada:2003bs} comes from the equation of motion
$\frac{\delta{\cal L}}{\delta A_i^-}-\d_+\frac{\delta{\cal
L}}{\delta (\d_+A_i^-)}=0$, where $A_i^-$ is the ``--" component of
the gauge field $A^{\mu}_i$ residing at the $i$-th site. However,
this equation of motion has nothing to do with the terms involving
the difference between fermions at different sites, which are the
cause of the doubling. Hence, even if we made some change(s) in the
standard Lagrangian e.g. by adding a Wilson term to fix the doubling
problem, we would not see any change in the equation of motion which
leads to the physical constraint.

It seems that SUSY algebra by itself resolves the species doubling
problem. This is indeed expected since we do not have any doubling
problems in boson sector and SUSY requires that the number of
degrees of freedom be the same for bosons and fermions. In general
it is difficult to maintain exact SUSY on a lattice, but it appears
that if it is achieved, then it automatically solves the species
doubling problem.  Clearly SDLCQ  is  one of a class of promising
approaches in the attempt to put a SYM theory on a spatial lattice
\cite{Cohen:2003xe}.

\section*{Acknowledgments}
This work was supported in part by the U.S. Department of Energy. We would like to
acknowledge the Aspen Center for Physics where part of this would was completed.

\end{document}